# ABOUT MAXIMUM ENERGY ACQUIRED BY A CHARGED PARTICLE IN THE FIELD OF A PLANE ELECTROMAGNETIC WAVE

V.A. Balakirev, I.N. Onishchenko

*National Science Center "Kharkiv Institute of Physics and Technology",*
*Kharkiv, Ukraine*
*E-mail: vabalakirev@ukr.net*

The energy characteristics of a relativistic charged particle in the field of a plane electromagnetic wave of a given amplitude are studied. The dependence of the particle's energy on its phase coordinate is obtained. The maximum value of the particle's energy during acceleration, as well as the acceleration length and time, are determined. The transverse displacement of the particle over the maximum acceleration length is determined.

## INTRODUCTION

With the advent of high-power lasers, the problem of accelerating charged particles in a vacuum using intense electromagnetic radiation has become increasingly relevant. To estimate the acceleration efficiency within this approach, one can use a simple model of the motion of a charged particle in the field of a plane electromagnetic wave. This problem admits an exact solution [1-10], which has a parametric form. In [9], energy relations for a charged particle in the field of a plane electromagnetic wave were obtained by averaging over the period of the high-frequency wave. Meanwhile, an ultrarelativistic particle can for a long time (over large distances) maintain synchronism between the particle and the wave, the phase velocity of which is equal to the speed of light. The time of synchronous interaction can be estimated as the time during which the phase shift of a particle at its movement is equal to $\pi$

$$t_{syn} \approx \gamma_0^2 T / 4 \gg T,$$

where $T$ is the period of the high-frequency wave, and $\gamma_0$ is the particle's relativistic factor. In this case, the energy relations of a particle in the field of an electromagnetic wave can be obtained by solving the problem exactly. Below, we study the question of the maximum energy acquired by a particle in the field of a plane electromagnetic wave using namely such approach.

## 1. STATEMENT OF THE PROBLEM. SOLUTION OF THE MOTION EQUATIONS

In the field of a plane electromagnetic wave with linear polarization:

electric field is $\vec{E} = \vec{A}\cos(\vec{k}\vec{r} - \omega t)$,

magnetic field is $\vec{H} = \dfrac{1}{k_0}\left[\vec{k} \times \vec{A}\right]\cos(\vec{k}\vec{r} - \omega t)$

a particle with a negative charge moves, $(\vec{k}\vec{A}) = 0$, where $\vec{A}$ is the wave amplitude, $\vec{k}$ is the wave vector, $\omega$ is the frequency, $k_0 = \omega/c$. The dimensionless relativistic equations of motion of a particle in the field of such a wave have the form

$$\frac{d\vec{p}}{dt} = -\left(\vec{a} + \left[\vec{v} \times \left[\vec{k} \times \vec{a}\right]\right]\right)\cos\vartheta, \quad \vartheta = t - \vec{k}\vec{r}, \tag{1}$$

$$\frac{d\vartheta}{dt} = \Omega, \tag{2}$$

$$\frac{d\vec{r}}{dt} = \frac{\vec{p}}{\gamma}, \tag{3}$$

where $\Omega = 1 - \vec{k}\vec{v}$. The equations of motion (1) - (3) are written in dimensionless variables:

$$\vec{p} \to \vec{p}/mc, \vec{v} \to \vec{v}/c, \vec{r} \to k_0 \vec{r}, t \to \omega t, \vec{v} = \frac{\vec{p}}{\gamma},$$

$\gamma = \sqrt{1 + \vec{p}^2}$ is the relativistic factor of the particle, $\vec{a} = \dfrac{e\vec{A}}{mc\omega}$ is the dimensionless amplitude, $a$ is the wave strength parameter. The initial conditions are arbitrary. Equation (1) can be represented as

$$\frac{d\vec{p}}{dt} = -\left[\Omega\vec{a} + (\vec{a}\vec{v})\vec{k}\right]\cos\vartheta. \tag{4}$$

Let us also give the equation for the particle energy

$$\frac{d\gamma}{dt} = -\vec{a}\vec{v}\cos\vartheta. \tag{5}$$

The system of equations (4), (2) has the vector integral of motion

$$\vec{k}\gamma - \vec{p} = \vec{a}\sin\vartheta + \vec{C}, \tag{6}$$

$$\vec{C} = \vec{k}\gamma_0 - \vec{p}_0 - \vec{a}\sin\vartheta_0 \tag{7}$$

is constant vector determined from the initial conditions, $2\pi > \vartheta_0 \geq 0$. The index "0" in (7) indicates the initial value of the corresponding quantity. We multiply scalarly the vector integral (6) by the vectors $\vec{k}$ and $\vec{a}$, respectively. As a result, instead of the vector integral, we obtain two scalar integrals

$$\gamma - \vec{k}\vec{p} \equiv \gamma\Omega = \vec{k}\vec{C} = \gamma_0 - \vec{k}\vec{p}_0 \equiv \gamma_0\Omega_0, \tag{8}$$

$$\vec{s}\vec{p} + a\sin\vartheta = -\vec{s}\vec{C} \equiv \vec{s}\vec{p}_0 + a\sin\vartheta_0, \tag{9}$$

where $\Omega_0 = 1 - \vec{k}\vec{v}_0$, $\vec{s} = \vec{a}/a$ is a unit vector in the direction of the electric field of the wave. From the vector integral (6) it follows also the relation

$$2\gamma\gamma_0\Omega_0 = 1 + \left[\vec{k}\gamma_0 - \vec{p}_0 - \vec{s}a(\sin\vartheta - \sin\vartheta_0)\right]^2, \tag{10}$$

which relates the relativistic factor of a particle to its phase coordinate. Taking into account integrals (8), (9), equations (5), (2) can be reduced to a system of coupled first-order nonlinear equations for the relativistic factor and phase coordinate

$$\frac{d\gamma}{dt} = \frac{a^2}{\gamma}(\sin\vartheta - \sin\vartheta_0 - \mu\vec{s}\vec{v}_0)\cos\vartheta, \tag{11}$$

$$\frac{d\vartheta}{dt} = \Omega_0\frac{\gamma_0}{\gamma}, \tag{12}$$

where $\mu = \gamma_0 / a$. From equation (12) it follows directly that always $d\vartheta / dt > 0$. Therefore, the particle phase only increases, respectively $\vartheta \geq \vartheta_0$. Of course, this system of equations contains integral (10). From this integral follows the following expression for the relativistic factor

$$\gamma = \gamma_0 + \frac{a^2}{2\gamma_0 \Omega_0}\left[(\sin\vartheta - \sin\vartheta_0)^2 - 2\mu\vec{s}\vec{v}_0(\sin\vartheta - \sin\vartheta_0)\right]. \quad (13)$$

The dependence of the particle's relativistic factor (i.e. energy) on the phase coordinate $\vartheta$ is periodic, and its value is limited. From relation (13) it follows the expression for the average over the phase coordinate

$$\bar{\gamma} = \frac{1}{2\pi}\int_0^{2\pi}\gamma(\vartheta)d\vartheta$$

the relativistic factor value

$$\bar{\gamma} = \gamma_0 + \Gamma_0(\vartheta_0), \quad (14)$$

where

$$\Gamma(\vartheta_0) = \frac{a^2}{2\gamma_0\Omega_0}\left(\frac{1}{2} + \sin^2\vartheta_0 + 2\mu\vec{s}\vec{v}_0 \sin\vartheta_0\right).$$

The implicit dependence of the phase coordinate on time with taking into account relation (12), follows from equation (11) and has the form

$$[\gamma_0 + \Gamma(\vartheta_0)](\vartheta - \vartheta_0) + \frac{a^2}{2\gamma_0\Omega_0}\left[2\sigma(\vartheta_0)(\cos\vartheta - \cos\vartheta_0) - \frac{1}{4}(\sin 2\vartheta - \sin 2\vartheta_0)\right] =$$
$$= \gamma_0\Omega_0 t. \quad (15)$$

Here $\sigma(\vartheta_0) = \sin\vartheta_0 + \mu\vec{s}\vec{v}_0$.

The vector equation for the radius-vector $\vec{r}(t)$ (3) is equivalent to two scalar equations for the displacements of the particle in two mutually orthogonal directions

$$\frac{d\vec{k}\vec{r}}{dt} = \frac{\vec{k}\vec{p}}{\gamma}, \quad \frac{d\vec{s}\vec{r}}{dt} = \frac{\vec{s}\vec{p}}{\gamma}. \quad (16)$$

In equations (16), it is convenient to use the phase coordinate $\vartheta(t)$ as the independent variable, rather than time $t$. Then, instead of equations (16), we obtain

$$\frac{d\vec{k}\vec{r}}{d\vartheta} = \frac{\gamma}{\gamma_0\Omega_0} - 1, \quad \frac{d\vec{s}\vec{r}}{d\vartheta} = \frac{a}{\gamma_0\Omega_0}(\mu\vec{s}\vec{v}_0 - \sin\vartheta + \sin\vartheta_0). \quad (17)$$

Solutions of these equations are easy to find

$$\vec{k}\vec{r} = \vec{k}\vec{r}_0 + \frac{1}{\gamma_0\Omega_0}\left\{[\vec{k}\vec{p}_0 + \Gamma(\vartheta_0)](\vartheta - \vartheta_0) + \frac{a^2}{2\gamma_0\Omega_0}\left[2\sigma(\vartheta_0)(\cos\vartheta - \cos\vartheta_0) - \frac{1}{4}(\sin 2\vartheta - \sin 2\vartheta_0)\right]\right\}, \quad (18)$$

$$\vec{s}\vec{r} = \vec{s}\vec{r}_0 + \frac{a}{\gamma_0\Omega_0}\left[\sigma(\vartheta_0)(\vartheta - \vartheta_0) + \cos\vartheta - \cos\vartheta_0\right], \quad (19)$$

$\vec{r}_0$ is radius-vector of the initial position of the particle. Relations (18), (19) determine the trajectory of the particle in space.

## 2. MAXIMUM ENERGY OF A PARTICLE

Let us investigate the dependence of the relativistic factor of a particle on the phase coordinate (13). The extreme points of this dependence are found from the condition

$$\frac{d\gamma}{d\vartheta} = \frac{a^2}{\gamma_0 \Omega_0} \cos\vartheta \left[\sin\vartheta - \sigma(\vartheta_0)\right] = 0. \tag{20}$$

And the nature of these points (maximum or minimum) is determined by the sign of the second derivative

$$\frac{d^2\gamma}{d\vartheta^2} = \frac{a^2}{\gamma_0 \Omega_0}\left[\cos 2\vartheta + \sigma(\vartheta_0)\sin\vartheta\right]. \tag{21}$$

From equation (20) we find the values of the phase coordinates of the extreme points

$$\vartheta_{ext} = \frac{\pi}{2} + \pi n, \tag{22}$$

where $n$ is an integer. In addition, if the condition

$$|\sigma(\vartheta_0)| < 1 \tag{23}$$

is satisfied for all values of the initial phases $\vartheta_0$, one more set of extreme points arises, which are found from the equation

$$\sin\vartheta = \sigma(\vartheta_0). \tag{24}$$

A necessary condition for the fulfillment of inequality (24) is the requirement

$$\mu \vec{s}\vec{v}_0 = \mu v_0 \cos\psi < 1,$$

$\psi$ is the angle between the vectors $\vec{v}_0$ and $\vec{s}$. Note that the extreme points (24) correspond to the equality $\vec{a}\vec{v} = 0$ in equation (5) for the relativistic factor.

In the interval of initial phases $2\pi > \vartheta_0 \geq 0$ on the period $2\pi > \vartheta \geq 0$, there are two extreme points

$$\vartheta_{ext} = \pi/2;\ 3\pi/2. \tag{25}$$

and the extreme points are arranged in the following order

$$\begin{cases} \pi/2 > \vartheta_0 \geq 0, \\ \qquad\qquad\qquad \vartheta_{ext} = \pi/2;\ 3\pi/2; \\ 2\pi > \vartheta_0 \geq 3\pi/2, \end{cases}$$

$$3\pi/2 > \vartheta_0 \geq \pi/2;\quad \vartheta_{ext} = 3\pi/2;\ \pi/2.$$

The second derivative of the relativistic factor with respect to the phase coordinate at the extreme points (25) is equal

$$\gamma''(\vartheta_{ext} = \pi/2) = -\frac{a^2}{\gamma_0 \Omega_0}\left[1 - \sigma(\vartheta_0)\right], \tag{26}$$

$$\gamma''(\vartheta_{ext} = 3\pi/2) = -\frac{a^2}{\gamma_0 \Omega_0}\left[1 + \sigma(\vartheta_0)\right]. \tag{27}$$

Let us firstly consider the case when condition (23) is satisfied. In this case, in addition to the extreme points (25), on the period $2\pi > \vartheta > 0$ there are two more extreme points

$$\vartheta_{ext} = \delta; \pi - \delta, \qquad (28)$$

$\delta = \arcsin\sigma(\vartheta_0)$ which are the roots of equation (24). The values of the second derivative at the extreme points (28)

$$\gamma''(\vartheta_{ext} = \delta, \pi - \delta) = \frac{a^2}{\gamma_0 \Omega_0}\left[1 - \sigma^2(\vartheta_0)\right]$$

are positive, and at the extreme points (25) its values (26), (27) are negative.

Thus, when condition (23) is satisfied, the dependence of the relativistic factor of a particle on the phase coordinate over the period $2\pi > \vartheta > 0$ has two alternating maxima and minima. The maxima are located at points (25), and the minima at points (28). The maximum and minimum values of the relativistic factor are equal in this case

$$\gamma_{max}(\vartheta = 3\pi/2) = \gamma_0 + \frac{a^2}{2\gamma_0 \Omega_0}(1 + \sin\vartheta_0)(1 + \sin\vartheta_0 + 2\mu\vec{s}\vec{v}_0), \qquad (29)$$

$$\gamma_{max}(\vartheta = \pi/2) = \gamma_0 + \frac{a^2}{2\gamma_0 \Omega_0}(1 - \sin\vartheta_0)(1 - \sin\vartheta_0 - 2\mu\vec{s}\vec{v}_0), \qquad (30)$$

$$\gamma_{min}(\vartheta = \delta; \pi - \delta) = \gamma_0 - \gamma_0\frac{(\vec{s}\vec{v}_0)^2}{2\Omega_0}. \qquad (31)$$

Note that the minimum value of the relativistic factor (31) does not depend on the wave amplitude and the initial value of the phase coordinate. The maximum values of the relativistic factor (29), (30), in turn, depend on the initial phase coordinate $\vartheta_0$. An analysis of functions (29), (30) for the extremum relative to the initial phase shows that the relativistic factor reaches its limiting values at $\vartheta_0 = \pi/2$ in the points $\vartheta = 3\pi/2 + 2\pi n$

$$\gamma_{lim}^{(+)} = \gamma_0 + \frac{2a^2}{\gamma_0\Omega_0}(1 + \mu\vec{s}\vec{v}_0), \qquad (32)$$

and at $\vartheta_0 = 3\pi/2$ in points $\vartheta = 5\pi/2 + 2\pi n$

$$\gamma_{lim}^{(-)} = \gamma_0 + \frac{2a^2}{\gamma_0\Omega_0}(1 - \mu\vec{s}\vec{v}_0). \qquad (33)$$

If $\vec{s}\vec{v}_0 > 0$, then the projection of the initial velocity vector onto the direction of electric field of the wave, taking into account the initial phase value $\vartheta_0 = \pi/2$, is negative, and the particle immediately enters in accelerating phase of the wave. In this case, the particle acquires the maximum possible energy (32). The particle with the initial phase $\vartheta_0 = 3\pi/2$ begins the interaction with the wave with deceleration. As a result, its maximum energy (33) is lower. If $\vec{s}\vec{v}_0 < 0$, then the opposite occurs. The particle with the initial phase $\vartheta_0 = 3\pi/2$ acquires higher energy.

The particle achieves its maximum energy values (32), (33) at the longitudinal $\vec{k}\vec{r}_{max}$ and transverse $\vec{s}\vec{r}_{max}$ lengths at time $t_{max}$

$$\vartheta_0 = \pi/2; \qquad \vec{k}(\vec{r}_{max}-\vec{r}_0) = \frac{\pi}{\gamma_0\Omega_0}\left[\vec{k}\vec{p}_0 + \frac{a^2}{2\gamma_0\Omega_0}\left(\frac{3}{2}+2\mu\vec{s}\vec{v}_0\right)\right],$$

$$\vec{s}(\vec{r}_{max}-\vec{r}_0) = \frac{\pi a}{\gamma_0\Omega_0}(1+\mu\vec{s}\vec{v}_0),$$

$$t_{max} = \frac{\pi}{\gamma_0\Omega_0}\left[\gamma_0 + \frac{a^2}{2\gamma_0\Omega_0}\left(\frac{3}{2}+2\mu\vec{s}\vec{v}_0\right)\right],$$

$$\vartheta_0 = 3\pi/2; \qquad \vec{k}(\vec{r}_{max}-\vec{r}_0) = \frac{\pi}{\gamma_0\Omega_0}\left[\vec{k}\vec{p}_0 + \frac{a^2}{2\gamma_0\Omega_0}\left(\frac{3}{2}-2\mu\vec{s}\vec{v}_0\right)\right],$$

$$\vec{s}(\vec{r}_{max}-\vec{r}_0) = -\frac{\pi a}{\gamma_0\Omega_0}(1-\mu\vec{s}\vec{v}_0),$$

$$t_{max} = \frac{\pi}{\gamma_0\Omega_0}\left[\gamma_0 + \frac{a^2}{2\gamma_0\Omega_0}\left(\frac{3}{2}-2\mu\vec{s}\vec{v}_0\right)\right].$$

The equations and relations obtained above are invariant with respect to the choice of coordinate system. We direct the longitudinal axis $z$ along the direction of wave propagation (along the vector $\vec{k}$) and the orthogonal axis $x$ along the electric field of a plane electromagnetic wave. In this coordinate system we have $z=\vec{k}\vec{r}$, $x=\vec{s}\vec{r}$, $\vec{k}\vec{p}_0 = p_{0z} = p\cos\varphi$, $\vec{s}\vec{p}_0 = p_{0x} = p\cos\psi$, where $\varphi$ is the angle between the vector $\vec{k}$ and the longitudinal axis, moreover $\psi = \pi/2 \mp \varphi$. The sign "−" is taken when $p_{0x} > 0$, and the sign "+" when $p_{0x} < 0$. At the same time $\cos\psi = \pm\sin\varphi$. Accordingly, the expression for the maximum value of the relativistic factor (32) can be written as follows

$$\gamma_{lim}^{(+)} = \gamma_0 + \gamma_0\frac{2}{\mu^2}F(\varphi). \qquad (34)$$

The function

$$F(\varphi) = \frac{1+\mu v_0\sin\varphi}{1-v_0\cos\varphi} \qquad (35)$$

describes the dependence of the maximum relativistic factor on the angle the electromagnetic wave propagation $\varphi \geq 0$. Note that function (32) is valid for the initial phase $\vartheta_0 = \pi/2$ at $p_{0x} < 0$ and the initial phase $\vartheta_0 = 3\pi/2$ when $p_{0x} > 0$. Analysis shows that function (35) reaches a maximum for the angle

$$\varphi_{max} = \arcsin u_{max}, \qquad (36)$$

where

$$u_{max} = \frac{1}{\gamma_0\left(\sqrt{a^2+1}+v_0 a\right)}. \qquad (37)$$

Note that for all values of the parameters included in (36) the inequality $u_{max} < 1$ is satisfied. The maximum value of the relativistic factor (34) for angle (35) is equal to

$$\gamma_{lim}^{(+)} = \gamma_0 + 2a\gamma_0\left(a + v_0\sqrt{a^2+1}\right). \qquad (38)$$

When $a^2 \gg 1$ and $v_0 = 1$ formulas (36) - (38) are simplified

$$\varphi_{max} = 1/2\gamma_0 a \ll 1,$$
$$\gamma_{lim}^{(+)} = \gamma_0 + 4\gamma_0 a^2. \tag{39}$$

## 3. LONGITUDINAL INJECTION OF A PARTICLE IN A WAVE FIELD

Let us consider separately the most interesting special case of particle injection strictly along the propagation of an electromagnetic wave (the longitudinal axis $z$). In this case, the general relations given above are significantly simplified. In particular, the dependence of the particle relativistic factor on the phase coordinate takes the form

$$\gamma = \gamma_0 + \frac{1+v_0}{2}\gamma_0 a^2 (\sin\vartheta - \sin\vartheta_0)^2, \tag{40}$$

$v_0$ is value of the longitudinal velocity of the particle. The trajectory of the particle on the plane $(x,z)$ is determined by parametric equations

$$x = x_0 + (1+v_0)\gamma_0 a\left[(\vartheta - \vartheta_0)\sin\vartheta_0 + \cos\vartheta - \cos\vartheta_0\right],$$

$$z = z_0 + (1+v_0)\gamma_0^2 \left\{\left[v_0 + \frac{1+v_0}{2}a^2\left(\frac{1}{2} + \sin^2\vartheta_0\right)\right](\vartheta - \vartheta_0) + \right.$$

$$\left. + \frac{1+v_0}{2}a^2\left[\sin\vartheta_0(\cos\vartheta - \cos\vartheta_0) - \frac{1}{4}(\sin 2\vartheta - \sin 2\vartheta_0)\right]\right\}.$$

The implicit dependence of a particle phase coordinate on time in the case under consideration follows from expression (15) and has the form

$$\left[1 + \frac{1}{2}(1+v_0)a^2\left(\frac{1}{2} + \sin^2\vartheta_0\right)\right](\vartheta - \vartheta_0) + (1+v_0)a^2\left[\sin\vartheta_0(\cos\vartheta - \cos\vartheta_0) - \right.$$

$$\left. - \frac{1}{8}(\sin 2\vartheta - \sin 2\vartheta_0)\right] = \frac{1+v_0}{\gamma_0^2}t.$$

The relativistic factor, determined by expression (40), reaches its maximum value for phase coordinate values (25)

$$\gamma_{max}(\vartheta = 3\pi/2) = \gamma_0 + \frac{1+v_0}{2}\gamma_0 a^2 (1+\sin\vartheta_0)^2,$$

$$\gamma_{max}(\vartheta = \pi/2) = \gamma_0 + \frac{1+v_0}{2}\gamma_0 a^2 (1-\sin\vartheta_0)^2.$$

The relativistic factor reaches its limiting value

$$\gamma_{lim} = \gamma_0 + 2(1+v_0)\gamma_0 a^2 \tag{41}$$

at the same initial phases $\vartheta_0 = \pi/2;\ 3\pi/2$. Note that formula (41) coincides with the previously obtained formula (38). The expression for the average value of the relativistic factor (14) in the case under consideration takes the form

$$\bar{\gamma} = \gamma_0 + \frac{3}{4}(1+v_0)\gamma_0 a^2.$$

The maximum energy gain $\gamma_{\lim} - \gamma_0$ is 8/3 times greater than the average energy gain $\bar{\gamma} - \gamma_0$. The distance at which the particle reaches its maximum value is

$$z_{\max} - z_0 = \pi(1+v_0)\gamma_0^2\left[v_0 + \frac{3(1+v_0)}{4}a^2\right].$$

At the same time the particle is deflected in the transverse direction by a distance

$$x_{\max} - x_0 = \pm\pi(1+v_0)\gamma_0 a.$$

The values of the specified coordinates are achieved at the moment of time

$$t_{\max} = \pi(1+v_0)\gamma_0^2\left[1 + \frac{3(1+v_0)}{4}a^2\right].$$

The longitudinal length and acceleration time are proportional to the square of the relativistic factor, and the transverse displacement is proportional to the relativistic factor raised to the first power.

## CONCLUSIONS

The acceleration of a relativistic charged particle by a plane electromagnetic wave is investigated. The dependence of the particle energy on its phase coordinate is obtained. The maximum value of the particle energy during acceleration, as well as the acceleration length and time, are determined. The transverse displacement of the particle along the maximum acceleration length is also determined. The most interesting case from the standpoint of experimental implementation of particle injection along the propagation of an electromagnetic wave is considered. It is shown that the maximum energy gain is proportional to the initial value of the relativistic factor and the square of the wave strength parameter. The length at which the particle acquires maximum energy is proportional to the square of the relativistic factor and the square of the wave strength parameter. Moreover, the transverse displacement is proportional to the relativistic factor and the wave strength parameter.